\documentclass[aps,prb,twocolumn,amssymb,showpacs,amsmath,superscriptaddress,floatfix]{revtex4}
\usepackage{graphicx}
\usepackage{bm}
\usepackage{multirow}

\begin{document}
\title {Electronic structure and spin-lattice relaxation in superconducting vortex
states on the kagome lattice near van Hove filling}

\author{Hong-Min Jiang}
\affiliation{School of Science, Zhejiang University of Science and
Technology, Hangzhou 310023, China}
\author{Shun-Li Yu}
\affiliation{National Laboratory of Solid State Microstructures and
Department of Physics, Nanjing University, Nanjing 210093, China}
\author{Xiao-Yin Pan}
\affiliation{Department of Physics, Ningbo University, Ningbo
315211,China}

\date{\today}

\begin{abstract}
Starting from a tight-binding model on the kagome lattice near the
van Hove filling, the superconducting (SC) properties are
investigated self-consistently using the Bogoliubov-de Gennes
equation with the consideration of the inequivalent third-neighbor
(TN) bonds. Near the van Hove filling, the most favorable SC
pairings are found to derive from the electrons belonging to the
same sublattice sites, including the on-site $s$-wave and the
spin-singlet/triplet TN pairings. The inequivalent TN bonds will
result in multiple SC components with different orbital angular
momentums (OAM) for the TN SC pairings. While the density of states
(DOS) and the temperature ($T$) dependence of the spin-lattice
relaxation rate ($T^{-1}_{1}$) exhibit distinct line shapes in the
SC state for the three cases, a peak structure in the $T$ dependence
of $T^{-1}_{1}$ can be found for both cases just below $T_{c}$ as a
result of the van Hove singularity, even though the SC gap has
nodes. The effects of magnetic vortices on the low energy
excitations and on the $T$ dependence of $T^{-1}_{1}$ with the
implications of the results are also discussed for both cases.
\end{abstract}
\maketitle

Recently, much attention has been focused on superconductivity in a
family of compounds AV$_{3}$Sb$_{5}$ (A=K, Rb, and
Cs)~\cite{Ortiz1,Ortiz2,QYin1,KYChen1,YWang1,ZZhang1,YXJiang1,
FHYu1,XChen1,HZhao1,HChen1,HSXu1,Liang1,CMu1,CCZhao1,SNi1,WDuan1,
Xiang1,PhysRevLett.126.247001,PhysRevX.11.041030,PhysRevB.104.L041101,
PhysRevX.11.041010,NatPhys.M.Kang,YFu1,HTan1,Shumiya1,FHYu2,LYin1,Nakayama2,
LNie1,HLuo1,Neupert1,YSong1,Nakayama1,HLi1}, which share a common
lattice structure with kagome net of vanadium atoms. Materials based
on kagome lattices have been predicted to host exotic quantum
physics because they embrace the geometrical lattice frustration,
the flat electronic bands, the Dirac cones and the topologically
nontrivial surface states. Meanwhile, the SC phase appears next to a
charge density wave phase in the pressure-temperature phase diagram.
As the electrons in these materials suffer simultaneously from the
geometrical frustration, topological band structure and the
competition between different possible ground states, the
observations of the superconductivity in these topological metals
are in themselves exotic and rare. The connection to the underlying
lattice geometry and the topological nature of the band structure
further places them in the context of wider research efforts in
topological physics and superconductivity.

To understand the underlying mechanism of the superconductivity in
kagome superconductors and its connection to the lattice geometry
and the topological nature of the band structure, numerous
experiments with various means were conducted in the past two years.
However, the inconsistent or even contradicting results were found
so far in experimental measurements and data analysis. The
temperature dependence of the nuclear spin-lattice relaxation rate
shows a Hebel-Slichter coherence peak just below $T_{c}$, indicating
that CsV$_{3}$Sb$_{5}$ is a nodeless $s$-wave
superconductor~\cite{CMu1}. The penetration depth measurements also
claim a nodeless gap~\cite{WDuan1}. Nevertheless, recent
measurements of thermal conductivity on CsV$_{3}$Sb$_{5}$ at
ultra-low temperature evidenced a finite residual linear term,
pointing to an unconventional nodal SC~\cite{CCZhao1}. In accordance
with this, the V-shaped SC gaps with residual zero-energy density of
states also suggest an anisotropic SC gap with
nodes~\cite{HSXu1,Liang1,HChen1}. Moreover, the STM experiment on
CsV$_{3}$Sb$_{5}$ at ultra-low temperature revealed a two-gap
structure with multiple sets of coherent peaks and residual
zero-energy DOS, accompanied by the magnetic/non-magnetic impurity
effect, implying a rather novel and interesting SC gap, i.e., the
sign preserved multiband superconductivity with gap
nodes~\cite{HSXu1}.

On the theoretical side, the vicinity to the van Hove filling was
proposed to be crucial to the superconductivity on the kagome
lattice. By using the variational cluster approach, the chiral
$d_{x^{2}-y^{2}}+id_{xy}$-wave SC state was found to be the most
favorable within a reasonable parameter range for the van Hove
filling kagome system based on the single-orbital Hubbard model with
the $1/6$ hole doping~\cite{SLYu1}. Moreover, in
Ref.~\onlinecite{SLYu1}, the sublattice character of the Bloch state
on the Fermi surface (FS) was shown to play a vital role in
determining the superconductivity of the kagome system, which was
also emphasized in the subsequent functional renormalization group
(FRG) studies~\cite{Kies1,WSWang1,Kies2}. By considering the
extended short-range interactions, the FRG studies on kagome systems
discovered a rich variety of electron instability, including
magnetism, charge order as well as superconductivity near  the van
Hove filling~\cite{Kies1,WSWang1,Kies2}. More recently, a random
phase approximation based on a two-orbital model revealed a $f$-wave
pairing instability over a large range of coupling strength,
succeeded by $d$-wave singlet pairing for stronger
coupling~\cite{XWu1}. Further more, it has been shown that the
coexistence of time-reversal symmetry breaking with a conventional
fully gapped superconductivity could lead to the gapless excitations
on the domains of the lattice symmetry breaking order~\cite{YGu1}.
The chiral flux phase has also been proposed to explain
time-reversal symmetry breaking in the kagome
superconductors~\cite{Feng1,Feng2}.

In view of the divergent experimental observations and the various
theoretical predictions, it is highly demanded to compare the
consequences of the theoretical predictions on the experimental
observations, especially with the emphases on the roles played by
the van Hove singularity and the inequivalent bonds on the kagome
lattice in a single-orbital Hubbard description. In this paper, we
carry out such an investigation on the SC pairing symmetries of the
kagome superconductors and compare their consequences on the
experimental observations. The study is to some extent an extension
to Ref.~\onlinecite{SLYu1} by incorporation of the three
inequivalent TN bonds on the kagome lattice. Starting from a
single-orbital tight-binding model on the kagome lattice near the
van Hove singularity at $1/6$ hole doping, the mean-field
calculations demonstrate that the most favorable SC pairings are
derived from the electrons belonging to the same sublattice sites,
including the on-site $s$-wave and the spin-singlet/triplet TNs
pairings, which are in line with the variational cluster
perturbation results. However, the incorporation of the inequivalent
TN bonds will lead to the SC pairing with multiple OAM components
with mixed $s_{ex}+(d\pm id')/(p\pm ip')+f$-wave symmetries, and
thus contributes to the two-gap structures of the DOS. Although the
spin-lattice relaxation exhibit distinct $T$ dependence for the
three cases, the Hebel-Slichter (or Hebel-Slichter-like) peak
structure can be found for both cases just below $T_{c}$ due to the
Fermi level being near the van Hove singularity. In the vortex
states, the cases for the on-site $s$-wave and the mixed
$s_{ex}+(d\pm id')$-wave parings possess discrete in-gap state
peaks, located on either side of the zero energy. Nevertheless, the
near-zero-energy in-gap state peak occurs in the vortex core for the
case of the mixed $(p\pm ip')+f$-wave paring. The vortices suppress
the Hebel-Slichter (or Hebel-Slichter-like) peaks of the
spin-lattice relaxation rate, but enhance them at low temperature.
While a sophisticated multi-orbital model is more appropriate to
make a direct comparison to the experimental results, the multiple
components of the SC pairing originated from the inequivalent bonds
on the kagome lattice and the Hebel-Slichter (or
Hebel-Slichter-like) peak of $T^{-1}_{1}$ below $T_{c}$ due the
vicinity to the van Hove filling are expected to persist in a
realistic multi-orbital description and should be reflected in the
experimental observations, provided that the system situates close
to the van Hove filling and the SC pairing is nonlocal.

\begin{figure}[htb]
\begin{center}
\vspace{-.2cm}
\includegraphics[width=230pt,height=196pt]{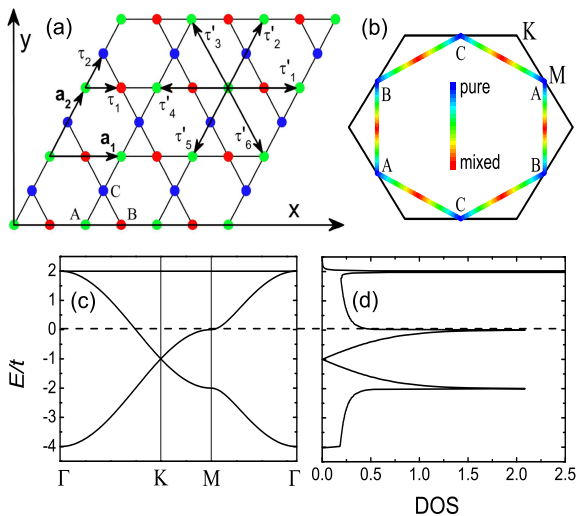}
\caption{(a) The lattice structure of the kagome superconductor,
made out of three sublattices $A$ (green dots), $B$ (red dots) and
$C$ (blue dots). $\mathbf{a_{1}}$ and $\mathbf{a_{2}}$ are two
translational vectors, $\tau_{1}$ and $\tau_{2}$ the
nearest-neighbor vectors, and $\tau'_{1}$-$\tau'_{6}$ the
third-neighbor vectors. (b) Fermi surface and weights of the
contribution to Fermi surface from three inequivalent lattice sites
$A$, $B$, and $C$ as represented by the colors. (c) The
tight-binding dispersion along high-symmetry cuts. The dashed line
is the Fermi level corresponding to the van Hove filling. (d) Normal
state density of states.}\label{fig1}
\end{center}
\end{figure}

The effective electron hoppings on a kagome lattice can be described
by the following tight-binding Hamiltonian,
\begin{eqnarray}
H_{0}&=&-\sum_{\langle
ij\rangle\sigma}(t_{ij}c^{\dag}_{i\sigma}c_{j\sigma}+h.c.)
-\mu\sum_{i\sigma}c^{\dag}_{i\sigma}c_{i\sigma},
\end{eqnarray}
where $c^{\dag}_{i\sigma}$ creates an electron with spin $\sigma$ on
the site $\mathbf{r}_{i}$ of the kagome lattice and $\langle
ij\rangle$ denotes nearest-neighbors (NN). $t_{ij}$ is the hopping
integral between the NN sites, and $\mu$ the chemical potential. For
the free hopping case with $t_{ij}=t$, the Hamiltonian $H_{0}$ can
be written in the momentum space,
\begin{eqnarray}
H_{0}(k)&=&\sum_{k\sigma}\Psi^{\dag}_{k\sigma}\mathcal{H}^{0}_{k}
\Psi_{k\sigma},
\end{eqnarray}
with $\Psi_{k\sigma}=(c_{Ak\sigma},c_{Bk\sigma},c_{Ck\sigma})^{T}$
and
\begin{eqnarray}
\mathcal{H}^{0}_{k}=-2t\left(
\begin{array}{ccc}
0 & \cos k_{1} &
\cos k_{2} \\
\cos k_{1} & 0 & \cos k_{3} \\
\cos k_{2} & \cos k_{3} & 0
\end{array}
\right).
\end{eqnarray}
The index $m=A,B,C$ in $c_{mk\sigma}$ labels the three basis sites
in the triangular unit cell. $k_{n}$ is abbreviated from
$\mathbf{k}\cdot\mathbf{\tau}_{n}$ with
$\mathbf{\tau}_{1}=\hat{x}/2$,
$\mathbf{\tau}_{2}=(\hat{x}+\sqrt{3}\hat{y})/4$ and
$\mathbf{\tau}_{3}=\mathbf{\tau}_{2}-\mathbf{\tau}_{1}$ denoting the
three NN vectors. The label of the sublattice sites, the NN vectors
and the translational vectors are shown in Fig.~\ref{fig1}(a). As
shown in Fig.~\ref{fig1}(c), the spectrum of $\mathcal{H}^{0}_{k}$
consists of one flat band $E^{(3)}_{k}=2t$ and two dispersive bands
\begin{eqnarray}
E^{(1,2)}_{k}&=&t(-1\pm\sqrt{4P_{k}-3}),
\end{eqnarray}
with $P_{k}=\cos^{2}k_{1}+\cos^{2}k_{2}+\cos^{2}k_{3}$. In addition
to the two inequivalent Dirac points formed by the touching points
of band 1 and 2 at $\mathbf{K}_{\pm}=(\pm2\pi/3,0)$ and the touching
point of band 2 and 3 at the center of the Brillouin zone (BZ),
there are three van Hove singularities with one originating from the
flat band, and the other two originating from the saddle points at
$M$ point of the BZ, as illustrated in Fig.~\ref{fig1}(d). The Fermi
levels at upper and lower saddle points correspond to the $1/2$ and
$1/6$ hole doping. Near the van Hove singularity at 1/6 hole doping,
the hexagonal FS shown in Fig.~\ref{fig1}(b) is similar to the ARPES
experimental observation and the DFT calculations~\cite{Ortiz1},
though a simple $d$-orbital tight-binding model was adopted. In the
calculations, we focus our study on the $1/6$ hole doping, as has
been down in Ref.~\onlinecite{SLYu1}.

The SC pairing is assumed to derive from the effective attractions
between electrons,
\begin{eqnarray}
H_{P}&=&V\sum_{ij,\sigma\sigma'}n_{i,\sigma}n_{j,\sigma'}.
\end{eqnarray}
In the mean-field approximation, the attractions can lead to the SC
pairings in the spin-singlet and spin-triplet channels respectively
as,
\begin{eqnarray}
H_{Ps}&=&\sum_{ij}(\Delta_{s,ij}c^{\dag}_{i,\uparrow}c^{\dag}_{j,\downarrow}
+h.c.),
\end{eqnarray}
and
\begin{eqnarray}
H_{Pt}&=&\sum_{ij}(\Delta_{t,ij}c^{\dag}_{i,\uparrow}c^{\dag}_{j,\downarrow}
+h.c.),
\end{eqnarray}
where the spin-singlet/triplet pairing potential is defined as
$\Delta_{s/t,ij}=\frac{V_{s/t}}{2}(\langle
c_{i,\uparrow}c_{j,\downarrow}\rangle\mp\langle
c_{i,\downarrow}c_{j,\uparrow}\rangle)$. Here, we consider the case
of spin-triplet pairing with the $\mathbf{d}$-vector along the
$z$-axis. Then, one obtains the total Hamiltonian as
\begin{eqnarray}
H&=&H_{0}+H_{Ps/t}.
\end{eqnarray}

Based on the Bogoliubov transformation, the diagonalization of the
Hamiltonian $H$ can be achieved by solving the following discrete
BdG equations,
\begin{eqnarray}
\sum_{j}\left(
\begin{array}{cc}
H_{ij,\sigma} &
\Delta_{s/t,ij} \\
\Delta^{\ast}_{s/t,ij} & -H^{\ast}_{ij,\bar{\sigma}}
\end{array}
\right)\left(
\begin{array}{cc}
u_{n,j,\sigma} \\
v_{n,j,\bar{\sigma}}
\end{array}
\right)= E_{n}\left(
\begin{array}{cc}
u_{n,i,\sigma} \\
v_{n,i,\bar{\sigma}}
\end{array}
\right),
\end{eqnarray}
where
$H_{ij,\sigma}=-t_{ij}\delta_{i+\mathbf{\tau}_{j},j}-\mu\delta_{i,j}$
with $\mathbf{\tau}_{j}$ denoting the four NN vectors.
$u_{n,i,\sigma}$ and $v_{n,i,\bar{\sigma}}$ are the Bogoliubov
quasiparticle amplitudes on the $i$-th site with corresponding
eigenvalue $E_{n}$. The SC pairing amplitude and electron densities
are obtained through the following self-consistent equations,
\begin{eqnarray}
\Delta_{s/t,ij}=&&\frac{V_{s/t}}{4}\sum_{n}(u_{n,i,\sigma}
v^{\ast}_{n,j,\bar{\sigma}}
\pm v^{\ast}_{n,i,\bar{\sigma}}u_{n,j,\sigma})\times \nonumber\\
&&\tanh(\frac{E_{n}}{2k_{B}T})
\nonumber\\
n_{i,\uparrow}=&&\sum_{n}|u_{n,i,\uparrow}|^{2}f(E_{n}) \nonumber\\
n_{i,\downarrow}=&&\sum_{n}|v_{n,i,\downarrow}|^{2}[1-f(E_{n})]..
\end{eqnarray}

\vspace*{.2cm}

At 1/6 hole doping, since the van Hove singularity at each saddle
point $M$ on the FS comes only from one of the three inequivalent
lattice sties, as shown in Fig.~\ref{fig1}(b), one can expect that
the favorable Cooper pairings are derived from two electrons
belonging to the same sublattice. Therefore, we consider the on-site
and the TN pairings, which are in the same sublattice. For the
kagome lattice, there are six TNs for each lattice site, and they
give rise to three inequivalent bonds, as denoted by $\tau'_{1}$,
$\tau'_{2}$ and $\tau'_{3}$ in Fig.~\ref{fig1}(a). While only
spin-singlet pairing is allowed for the on-site pairing, both the
spin-singlet and spin-triplet pairings are permissible for the TN
bonds. Since the three bonds are different, we only set the pairing
on each bond to be spin-singlet or spin-triplet, and let the pairing
amplitude on the three different bonds to be determined
self-consistently. For the TN pairings, we have
$\Delta_{s,\tau'_{4}}=\pm\Delta_{s,\tau'_{1}}$,
$\Delta_{s,\tau'_{5}}=\pm\Delta_{s,\tau'_{2}}$ and
$\Delta_{s,\tau'_{6}}=\pm\Delta_{s,\tau'_{3}}$ for the
spin-singlet/triplet pairings. In the calculations, we choose the
effective pairing interactions $V_{s}=V_{s0}=1.6$ for the on-site
$s$-wave pairing, $V_{s}=V_{s1}=1.2$ for the TN spin-singlet
pairing, and $V_{t}=1.4$ for the TN spin-triplet pairing
respectively to give rise to the comparable SC transition
temperatures for the three cases. At zero field, the self-consistent
results of the TN pairing amplitudes on the three different bonds
around three sublattice sites are displayed in table I.

\begin{table}[]
\begin{tabular}{|l|l|l|l|l|}
\hline
                                 &        & $\Delta_{s/t,\tau'_{1}}$ & $\Delta_{s/t,\tau'_{2}}$ & $\Delta_{s/t,\tau'_{3}}$ \\ \hline
 \multirow{3}{*}{TN spin-singlet} & A site & 0.05 & 0.05 & -0.03 \\ \cline{2-5}
                                  & B site & -0.03 & 0.05 & 0.05 \\ \cline{2-5}
                                  & C site & 0.05 & -0.03 & 0.05 \\ \hline
 \multirow{3}{*}{TN spin-triplet} & A site & -0.04 & 0.04 & -0.067 \\ \cline{2-5}
                                  & B site & -0.067 & 0.04 & -0.04 \\ \cline{2-5}
                                  & C site & -0.04 & 0.067 & -0.04 \\ \hline
\end{tabular}
\caption{Pairing strength on the three inequivalent TN bonds for the
three sublattice sites.}
\end{table}

The different pairing strengths on the three inequivalent bonds will
generally lead to a SC pairing with multiple components of the
orbital angular momentum. It would be useful to get some perspective
on the symmetries of the TN bond SC pairings in the kagome lattice
from the real space description. In real space, the pairing
amplitude on site $\mathbf{r}_{i}$ is generally defined as,
\begin{eqnarray}
\Delta^{l}_{s/t,+/-}(\mathbf{r}_{i})=&&\frac{1}{N_{c}}\sum_{\tau'_{j}}\Delta_{s/t,\tau'_{j}}e^{il\theta_{+/-}(\tau'_{j})}.
\end{eqnarray}
Here, $\Delta^{l}_{s/t,+/-}(\mathbf{r}_{i})$ stands for the
clockwise/anticlockwise ($+/-$) spin-singlet/triplet ($s/t$) pairing
with orbital momentum $l$ in unit of $\hbar$, which determines the
spatial symmetry of the Cooper pair wavefunction.
$\theta_{+/-}(\tau'_{j})$ denotes the polar angle of the TN bond
measured anticlockwise/clockwise from the $x$-axis, and $N_{c}$ is
the number of the TN sites around $\mathbf{r}_{i}$. In Eq. (11),
$\Delta^{l}_{s/t,+/-}(\mathbf{r}_{i})$ with $l=0,1,2,3...$ picks up
respectively the orbital components of $s,p,d,f...$-waves, and the
relationship between $\Delta^{l}_{s/t,+}(\mathbf{r}_{i})$ and
$\Delta^{l}_{s/t,-}(\mathbf{r}_{i})$ tells us the information about
the pairing chirality. From Eq. (11) and the self-consistent results
shown in table I, one could get mixed $s_{ex}+(d\pm id')$-wave
symmetry for the TN spin-singlet pairing and mixed $(p\pm
ip')+f$-wave symmetry for the TN spin-triplet pairing. In reciprocal
space, the components with different orbital momentums have the
following form,
\begin{eqnarray}
\Delta_{s_{ex}}(\mathbf{k})=&&\Delta^{s_{ex}}_{0}[\cos(k_{x})+2\cos(k_{x}/2)\cos(\sqrt{3}k_{y}/2)];
\nonumber\\
\Delta_{d\pm id'}(\mathbf{k})=&&\Delta^{d\pm
id'}_{0}[\cos(k_{x})-\cos(k_{x}/2)\cos(\sqrt{3}k_{y}/2)\nonumber\\
&&\pm i\sqrt{3}\sin(k_{x}/2)\sin(\sqrt{3}k_{y}/2)]; \nonumber\\
\Delta_{p\pm ip'}(\mathbf{k})=&&\Delta^{d\pm
id'}_{0}[\sin(k_{x})+\sin(k_{x}/2)\cos(\sqrt{3}k_{y}/2)\nonumber\\
&&\pm i\sqrt{3}\cos(k_{x}/2)\sin(\sqrt{3}k_{y}/2)]; \nonumber\\
\Delta_{f}(\mathbf{k})=&&\Delta^{f}_{0}[\sin(k_{x})-2\sin(k_{x}/2)\cos(\sqrt{3}k_{y}/2)],
\end{eqnarray}
where $\Delta^{s_{ex}}_{0}=|\Delta^{0}_{s}|$, $\Delta^{d\pm
id'}_{0}=|\Delta^{2}_{s,\pm}|$, $\Delta^{p\pm
ip'}_{0}=|\Delta^{1}_{t,\pm}|$ and
$\Delta^{f}_{0}=|\Delta^{3}_{t}|$. The values of the $s_{ex}/f$-wave
$\Delta^{s_{ex}}_{0}/\Delta^{f}_{0}$ and the $d\pm id'/p\pm
ip'$-wave components $\Delta^{d\pm id'}_{0}/\Delta^{p\pm ip'}_{0}$
are shown in table II, and the sign distributions of the pairing
components refer to Ref.~\onlinecite{SLYu1} for details. While the
$s_{ex}$- and $d\pm id'$-waves components exhibit comparable
strength for the TN spin-singlet pairing, the $f$-wave component
dominates over the $p\pm ip'$-wave component for the TN spin-triplet
pairing. The smallness of the $p\pm ip'$-wave component in the mixed
$(p\pm ip')+f$-wave symmetry pairing state will not remove the nodes
of the $f$-wave pairing but will move them, resulting in an unusual
SC pairing state with accidental nodes. The equality of
$\Delta^{d+id'/p+ip'}_{0}$ and $\Delta^{d-id'/p-ip'}_{0}$ dictates
the two degenerate SC paring states with right and left chiralities
in the $d\pm id'/p\pm ip'$-wave component at zero field.

The three typical SC parings in their uniform SC states produce
distinct site-averaged DOS spectra
$N(E)=\frac{1}{N}\sum_{i}N(E,\mathbf{r}_{i})$ with definition
$N(E,\mathbf{r}_{i})=N_{\uparrow}(E,\mathbf{r}_{i})+N_{\downarrow}(E,\mathbf{r}_{i})=-\sum_{n}[|u_{i,\uparrow}^{n}|^{2}f^{'}(E_{n}-E)
+|v_{i,\downarrow}^{n}|^{2} f^{'}(E_{n}+E)]$, which is proportional
to the differential tunneling conductance observed in scanning
tunneling microscopy (STM) experiments. The results are summarized
by the solid black lines at the bottom of each panels in
Fig.~\ref{fig3}. For the on-site $s$-wave symmetry, a single
U-shaped full gap structure can be seen in the DOS as shown by the
solid black line in Fig. 3(a), depicting a typical feature for the
isotropic SC gap without nodes along the FS. In the $s_{ex}+(d\pm
id')$-wave symmetry, the DOS consists of a small U-shaped gap
structure at low energy and a broad V-shaped gap structure at higher
energy as shown in Fig.~\ref{fig3}(b), presenting an anisotropic
nodeless two-gap structure. As for the $(p\pm ip')+f$-wave symmetry
shown in Fig.~\ref{fig3}(c), dominant $f$-wave component plus a tiny
value of $p\pm ip'$-wave component produce a broad V-shaped gap
structure inlaid by a small V-shaped gap with residual DOS at zero
energy in the SC state, displaying a characteristic of nodal two-gap
SC pairing. We note that the V-shaped SC gap with multiple sets of
coherent peaks and residual zero-energy DOS are in good accordance
with the STM experiments~\cite{HSXu1,Liang1,HChen1}.

\begin{table}[]
\begin{tabular}{|l|l|l|l|l|}
\hline
 & $\Delta^{s_{ex}/f}_{0}$ & $\Delta^{d+id'/p+ip'}_{0}$ & $\Delta^{d-id'/p-ip'}_{0}$ \\ \hline
 TN spin-singlet & 0.023 & 0.027 & 0.027 \\ \hline
 TN spin-triplet & 0.049 & 0.009 & 0.009 \\ \hline
\end{tabular}
\caption{Pairing amplitude for the different orbital components.}
\end{table}

Now we address the vortex structure of the three types of the SC
states. In the presence of a perpendicular magnetic field, the
hopping terms are described by the Peierls substitution. For the NN
hopping between sites $i$ and $i+\mathbf{\tau}_{j}$, one has
$t_{i,i+\mathbf{\tau}_{j}}=te^{i\varphi_{i,i+\mathbf{\tau}_{j}}}$,
where $\varphi_{i,i+\mathbf{\tau}_{j}(\mathbf{\tau}'_{j})}=
\frac{\pi}{\Phi_{0}}\int^{r_{i}}_{r_{i+\mathbf{\tau}_{j}(\mathbf{\tau}'_{j})}}\mathbf{A}(\mathbf{r})\cdot
d\mathbf{r}$ with $\Phi_{0}=\frac{hc}{2e}$ being the SC flux quanta.
In this case, the pairing amplitude on site $\mathbf{r}_{i}$ is
reformulated as
$\Delta^{l}_{s/t,L/R}(\mathbf{r}_{i})=\frac{1}{N_{c}}\sum_{\tau'_{j}}\Delta_{s/t,\tau'_{j}}e^{il\theta_{R/L}(\tau'_{j})}
e^{i\varphi_{i,i+\mathbf{\tau}'_{j}}}$. In the calculations, we
consider a parallelogram vortex unit cell with size of
$22\mathbf{a_{1}}\times44\mathbf{a_{2}}$ as shown in
Fig.~\ref{fig1}(a), where two vortices are accommodated. The vector
potential $\mathbf{A}(\mathbf{r})=(0,Bx,0)$ is chosen in the Landau
gauge to give rise to the magnetic field $\mathbf{B}$ along the
$z$-direction.

\vspace*{.2cm}
\begin{figure}[htb]
\begin{center}
\vspace{-.2cm}
\includegraphics[width=260pt,height=290pt]{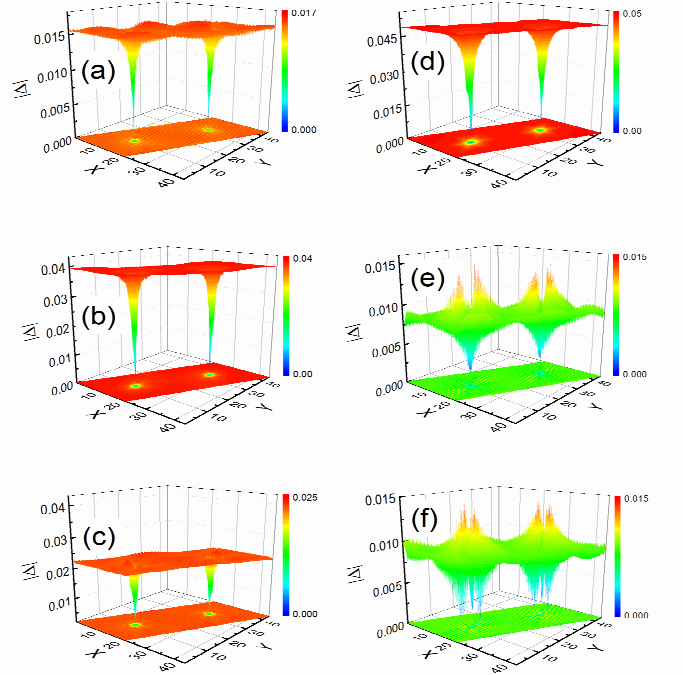}
\caption{The spatial distributions of the SC order parameters in the
vortex states for the mixed $s_{ex}+(d\pm id')$- (left panel), and
$(p\pm ip')+f$-wave (right panel) symmetries. (a), (b) and (c) show
the spatial distributions of the amplitudes for the $s_{ex}$,
$d+id'$ and $d-id'$ components, respectively. (d), (e) and (f) show
the spatial distributions of the amplitudes for the $f$, $p+ip'$ and
$p-ip'$ components, respectively. }\label{fig2}
\end{center}
\end{figure}

Under a perpendicular magnetic field, the vanishment of the
screening current density at the vortex center drives the system
into the vortex states with the suppression of the SC order
parameter around the vortex core, thereby forming a vortex with
winding $+1$. In the vortex states, the SC order parameter
$|\Delta({\mathbf{r}_{i}})|$ vanishes at the vortex core center and
recovers its bulk value at the core edge with the core size
$\xi_{1}$ on the scale of coherent length $\xi_{0}$, as can be seen
from Fig.~\ref{fig3} for both cases (The case of on-site $s$-wave
pairing is not shown here). Besides the standard SC vortex
structure, there are two prominent features to be specified in the
vortex states for the SC pairings with multiple OAM components on
the kagome lattice. Firstly, consistent with the STM experimental
observations in CsV$_{3}$Sb$_{5}$~\cite{HSXu1} and similar to the
observations in NbSe$_{2}$~\cite{Hess1,Haya1,Haya2} and
YNi$_{2}$B$_{2}$C~\cite{Nish1,Kane1,Naga1}, the vortex core has a
typical star shape with sixfold symmetry for both cases, reflecting
the underlying crystalline band structure. Secondly, the afore
mentioned two degenerate SC paring states with right and left
chiralities in the $d\pm id'/p\pm ip'$-wave component are removed
under a perpendicular magnetic field, because $d\pm id'/p\pm
ip'$-wave components correspond to states with an internal phase
winding of the Cooper pairs along the $z$-axis. In the mixed
$s_{ex}+(d\pm id')$-wave pairing state, the comparable strength for
the $s_{ex}$- and $d\pm id'$-wave components renders both of them to
response effectively to the magnetic field. The internal phase of
the $d+id'$-wave component has a $-2$ winding, which counteracts the
phase winding $+1$ of the vortex to save the energy cost of
supercurrents. As a result, the application of the magnetic field
transfers the weight from the $s_{ex}$- and $d-id'$-wave components
to the $d+id'$-wave component in the mixed $s_{ex}+(d\pm id')$-wave
pairing state, as evidenced by a comparison of table II with the
spatial distributions of the SC order parameters in
Figs.~\ref{fig2}(a),~\ref{fig2}(b) and~\ref{fig2}(c). On the other
hand, the screening current density from the dominant $f$-wave
component in the mixed $(p\pm ip')+f$-wave symmetry pairing
diminishes the impact of the magnetic field on the $p\pm ip'$-wave
components, so there is a little degeneracy lifting for the two
chiral $p\pm ip'$-wave components as shown in Figs.~\ref{fig2}(e)
and~\ref{fig2}(f), despite the $p-ip'$-wave component possessing the
internal phase winding $-1$.

Then, we pursuit the electronic structures in the vortex states by
examining the energy dependence of the LDOS. In order to reduce the
finite size effect, the calculations of the LDOS are carried out on
a periodic lattice which consists of $16\times 8$ parallelogram
supercells, with each supercell being the size
$22\mathbf{a_{1}}\times44\mathbf{a_{2}}$. In Fig.~\ref{fig3}, we
show the energy dependence of the LDOS on a series of sites along
the long side direction of the parallelogram moving away from the
core center. Since both the on-site $s$-wave and the mixed
$s_{ex}+(d\pm id')$-wave pairings are fully gaped, similar in-gap
states appear in the core region. At the vortex center, the
Caroli-de Gennes-Matricon states at the vortex center accumulate to
give rise to two peaks reside on each side about the zero energy,
forming a small gap at the zero energy. As the site moving away from
the vortex center, the two peaks depart further and fade away, as
presented in Figs.~\ref{fig3}(a) and~\ref{fig3}(b). For the mixed
$(p\pm ip')+f$-wave symmetry pairing, by contrast, a near-zero peak
appears at the vortex center, which does not disperse in a large
distance as moving away from the vortex center. It is worth noting
that the near-zero energy peak and the dispersionless of the peak
are again in excellent agreement with the STM experimental
observations~\cite{Liang1}.

\vspace*{.2cm}
\begin{figure}[htb]
\begin{center}
\vspace{-.2cm}
\includegraphics[width=240pt,height=130pt]{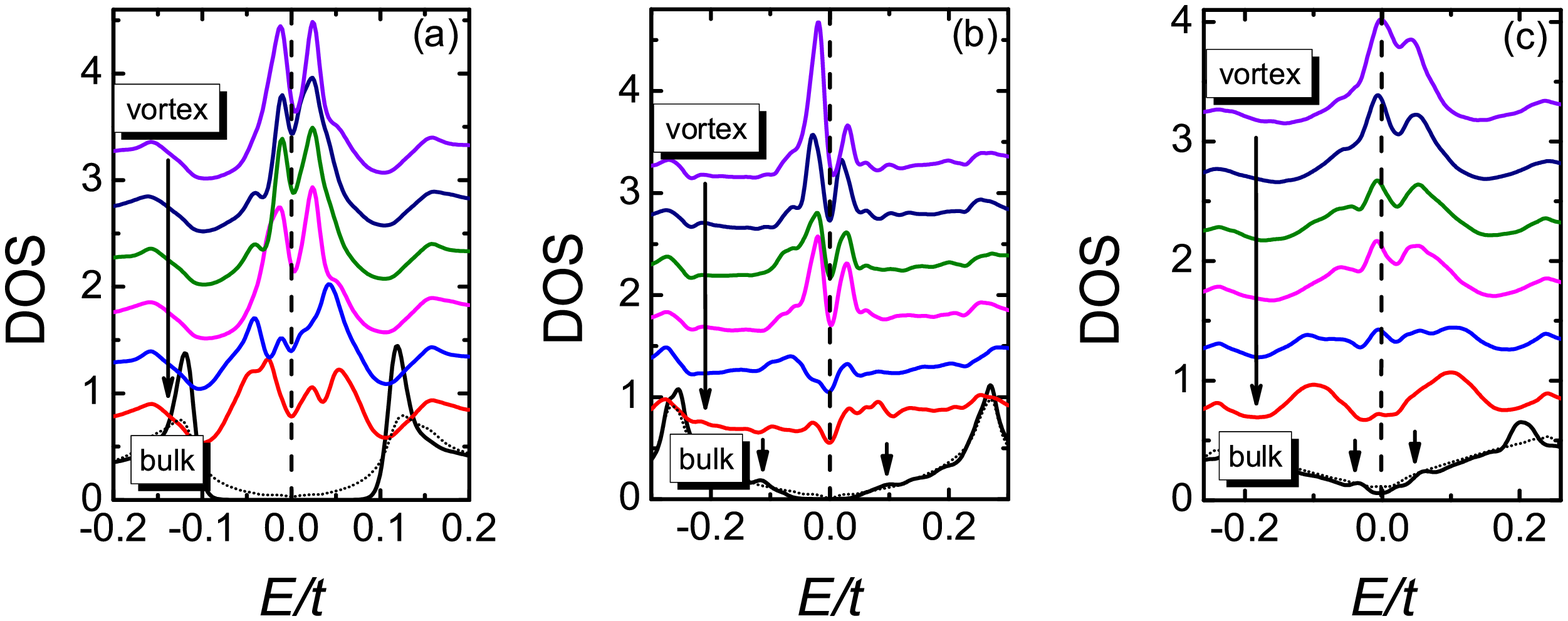}
\caption{The energy dependence of the LDOS on a series of sites for
on-site spin-singlet pairing (a), spin-singlet paring on the
same-sublattice bond (b), and spin-triplet paring on the
same-sublattice bond (c). In each panel from top to bottom, the
curves stand for the LDOS at sites along the long side direction of
the parallelogram moving away from the core center. The curves are
vertically shifted for clarity. At the bottom of each panels, the
DOS in the uniform SC state and the site-averaged DOS in the vortex
states are plotted as solid black lines and dotted black lines,
respectively. The dashed vertical lines in each panel denote the
position of the zero bias, and the short arrows in (b) and (c) mark
the secondary gap edges.}\label{fig3}
\end{center}
\end{figure}

Next, we turn to the discussion of the $T$ dependence of
$T^{-1}_{1}$. The site-dependent
$T^{-1}_{1}(\mathbf{r}_{i},\mathbf{r}_{i'})$ is given
by~\cite{Takigawa,Jiang1}
\begin{eqnarray}
R(\mathbf{r}_{i},\mathbf{r}_{i'})=&&\textmd{Im}\chi_{+,-}(\mathbf{r}_{i},\mathbf{r}_{i'},i\Omega_{n}\rightarrow\Omega+i\eta)/(\Omega/T)|_{\Omega\rightarrow
0}\nonumber\\
=&&-\sum_{n,n'}[u_{n,i}u^{\ast}_{n,i'}v_{n',i}v^{\ast}_{n',i'}
-v_{n,i}u^{\ast}_{n,i'}u_{n',i}v^{\ast}_{n',i'}]\nonumber\\
&&\times\pi T f'(E_{n})\delta(E_{n}-E_{n'}).
\end{eqnarray}
We choose $\textbf{r}_{i}=\textbf{r}_{i'}$ because the nuclear
spin-lattice relaxation at a local site is dominant. Then the
site-dependent relaxation time is given by
$T_{1}(\mathbf{r}_{i})=1/R(\mathbf{r}_{i},\mathbf{r}_{i})$ and the
bulk relaxation time $T_{1}=(1/N)\sum_{i}T_{1}(\mathbf{r}_{i})$. In
the calculations, we adopt
$\delta(E_{n}-E_{n'})=\pi^{-1}\textmd{Im}(E_{n}-E_{n'}-i\eta)$ with
typical value $\eta=0.01$. In a conventional $s$-wave
superconductor, the $T$ dependence of $T^{-1}_{1}$ develops a peak
structure below $T_{c}$, which is called Hebel-Slichter coherence
peak as observed experimentally in SC Al by Hebel and
Slichter~\cite{Hebel}, and explained theoretically as a result of
the enhancement of the SC DOS at the gap edge along with the
non-zero coherent factor described in BCS theory~\cite{Hebel}. Thus
the observation of the Hebel-Slichter peak below $T_{c}$ is usually
considered to be the hallmark for $s$-wave
superconductivity~\cite{CMu1}.

\begin{figure}[htb]
\begin{center}
\vspace{-.2cm}
\includegraphics[width=240pt,height=290pt]{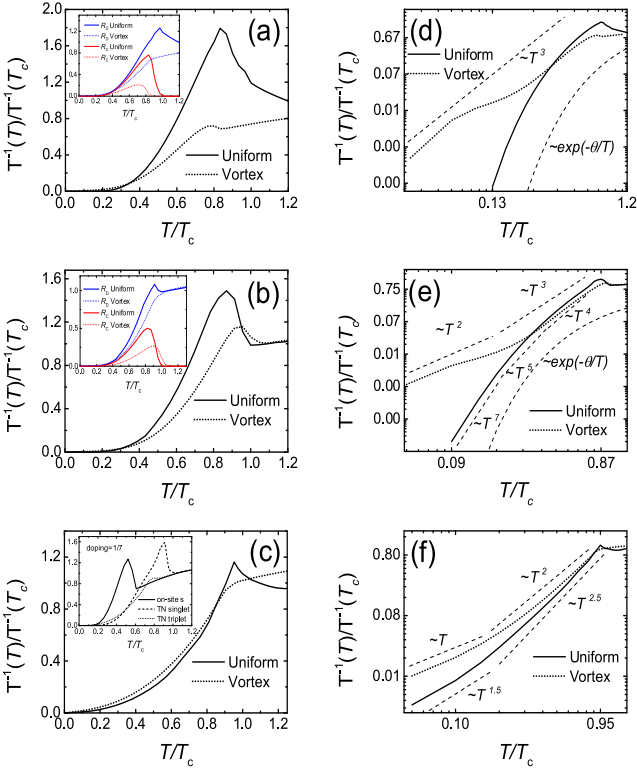}
\caption{Left panels: $T$-dependence of $\Delta(T)$ and
$T_{1}^{-1}$. Right panels: $T$-dependence of $T_{1}^{-1}$ shown in
the double logarithmic chart. (a) and (d) show the results for the
on-site pairing, (b) and (e) the results for the spin-singlet paring
on the TN bond, and (c) and (f) the results for the spin-triplet
paring on the TN bond. Insets in (a) and (b) display the $T$
evolutions of $R_{D}$ and $R_{C}$ (see text). Inset in (c) gives the
results of $T$ dependence of $T^{-1}_{1}$ for the three cases at
doping $1/7$.}\label{fig4}
\end{center}
\end{figure}

In the absence of the magnetic field, the Hebel-Slichter (or
Hebel-Slichter-like) peaks of $T^{-1}_{1}$ below $T_{c}$ are
evidenced in Figs.~\ref{fig4}(a)-(c) for both cases. It is quite
remarkable for the case of the mixed $(p\pm ip')+f$-wave with nodal
SC gap. To figure out the origin as well as the different nature of
the peaks, we show in the same figures the temperature evolution of
$R_{D}\equiv-\frac{1}{N}\sum_{i,n,n'}(u_{n,i}u^{\ast}_{n,i'}v_{n',i}v^{\ast}_{n',i'})\pi
T f'(E_{n})\delta(E_{n}-E_{n'})$ and
$R_{C}\equiv\frac{1}{N}\sum_{n,n'}(v_{n,i}u^{\ast}_{n,i'}u_{n',i}v^{\ast}_{n',i'})\pi
T f'(E_{n})\delta(E_{n}-E_{n'})$, i.e., the contributions from the
first and second terms in Eq. (12) to $T^{-1}_{1}$. $R_{D}$ is
proportional to $N_{\uparrow}\times N_{\downarrow}$, which gives
rise to the enhancement of the SC DOS at the gap edge with the
enhancement depending upon the sharpness of the SC gap edge and the
specific DOS of the normal state on where the SC gap opens. On the
other hand, $R_{C}$ describes the coherent effect of the SC state.
As is seen in Fig.~\ref{fig4}(c) and the insets of
Figs.~\ref{fig4}(a) and (b), the temperature evolutions of $R_{D}$
develop a peak just below $T_{c}$ for both cases (Note that only
$R_{D}$ contributes to $T^{-1}_{1}$, and accordingly
$R_{D}=T^{-1}_{1}$ for the mixed $(p\pm ip')+f$-wave pairing, as
will be shown in the following.), due to the fact that the
infinitesimal SC gaps opening at the van Hove singularity with
divergent DOS would also have divergent DOSs at the gap edges.
However, the $T$ dependence of $R_{C}$ is different for the three
cases. Specifically, $R_{C}$ evolves a peak below $T_{c}$ for the
cases of the on-site $s$- and the mixed $s_{ex}+(d\pm id')$-wave
symmetries, whereas it remains zero for the case of the mixed $(p\pm
ip')+f$-wave symmetry. This can be understood by noting that the
mixed triplet $(p\pm ip')+f$-wave pairing with odd parity
$\Delta_{ji}=-\Delta_{ij}$ forbids the local SC correlation
$u_{n,i}v^{\ast}_{n,i}$.

Thus far, we have demonstrated that the results for the mixed $(p\pm
ip')+f$-wave symmetry reconcile the various inconsistent or
apparently contradicting experiments, including the V-shaped SC gap
with residual DOS at zero energy, the dispersionless of the
near-zero energy peak in the vortex core, as well as the
Hebel-Slichter-like peak of the $T$ dependence of $T^{-1}_{1}$.
While the appearance of the Hebel-Slichter-like peak for the mixed
$(p\pm ip')+f$-wave symmetry here seems to support the NMR
experiment, its origin is different in nature from the
Hebel-Slichter coherent peak. The Hebel-Slichter coherent peak for
the on-site $s$- and the mixed $s_{ex}+(d\pm id')$-wave paring
symmetries derives from the simultaneous enhancement of $R_{D}$ and
$R_{c}$, but the peak for the case of the mixed $(p\pm ip')+f$-wave
symmetry originates merely from the enhancement of $R_{D}$. Due to
the nodal SC gap of the mixed $(p\pm ip')+f$-wave pairing, the
sharpness of the SC gap edge is weakened as the Fermi level
deviating from the van Hove singularity, and this in turn undermines
the Hebel-Slichter-like peak for the case of the mixed $(p\pm
ip')+f$-wave symmetry. As a result, the Hebel-Slichter-like peak for
the case of the mixed $(p\pm ip')+f$-wave symmetry diminishes and
eventually disappears with the Fermi level deviation from the van
Hove filling. This is verified in the inset of Fig.~\ref{fig4}(c)
for a specified doping level $1/7$. By contrast, the Hebel-Slichter
peak remains robust for the cases of the on-site $s$- and the mixed
$s_{ex}+(d\pm id')$-wave paring symmetries. To verify or falsify the
above scenario, the NMR experiments on different doping levels are
encouraged to observe the doping evolutions of the
Hebel-Slichter-like peak.

Below $T_{c}$, the three cases, however, exhibit distinct $T$
dependence of $T_{1}^{-1}$. The on-site $s$-wave pairing evolves
into an exponential dependence below $T_{c}$, as presented by the
solid line in Fig.~\ref{fig4}(d), which is the consequence of the
full-gaped DOS in Fig.~\ref{fig3}(a). The gap anisotropy of the
mixed $s_{ex}+(d\pm id')$-wave pairing changes the exponential
dependence to a power law relation $T_{1}^{-1}\sim T^{\alpha}$ with
$\alpha$ varying from $4$ to $5$ below $T_{c}$ and $T_{1}^{-1}\sim
T^{7}$ at low temperature, as displayed by the solid line in
Fig.~\ref{fig4}(e). For the case of $(p\pm ip')+f$-wave symmetry
pairing, the $T$ dependence of $T_{1}^{-1}$ changes its line-shape
further to $T_{1}^{-1}\sim T^{2.5}$ below $T_{c}$ and
$T_{1}^{-1}\sim T^{1.5}$ at low temperature, shown in
Fig.~\ref{fig4}(f), as a result of the V-shaped gap and the residual
DOS at zero energy.

In the presence of the perpendicular magnetic field, on one hand,
the intensity of the Hebel-Slichter (or Hebel-Slichter-like) peaks
are suppressed by localized excitations within the vortex
cores~\cite{Takigawa,Curro1}. While strong depression of the peak of
$R_{C}$ below $T_{c}$ can be seen for the case of the on-site
$s$-wave pairing [see the inset of Fig.~\ref{fig4}(a)], the
depression is just moderate with the peak position shifting slightly
toward higher temperature for the mixed $s_{ex}+(d\pm id')$-wave
paring symmetry [the inset of Fig.~\ref{fig4}(b)], owing to the
offsetting effect of the internal phase of the $d+id'$-wave
component [see Fig.~\ref{fig2}(b)]. As a result, one could barely
see a trail of the peak as shown by the dotted black line in
Fig.~\ref{fig4}(a) for the on-site $s$-wave pairing symmetry, but
still evidence a robust peak feature with its position moving
slightly to higher temperature for the mixed $s_{ex}+(d\pm
id')$-wave paring symmetry [refer to the dotted black line in
Fig.~\ref{fig4}(b)]. Nevertheless, the peak of $R_{D}$ below $T_{c}$
is suppressed completely for both cases, due to the blunting of the
gap edges as shown by the dotted black lines in
Figs.~\ref{fig4}(a)-(c). This directly leads to the disappearance of
the Hebel-Slichter-like peak for the case of the mixed $(p\pm
ip')+f$-wave symmetry, as presented by the dotted line in
Fig.~\ref{fig4}(c). On the other hand, the main effect of the
vortices is to enhance the $T$ dependence of $T^{-1}_{1}$ for all
symmetries at low temperature. The enhancement of $T^{-1}_{1}$ is
exemplified in Fig.~\ref{fig4}(a) by changing the exponential $T$
dependence to roughly $T^{3}$ below $T_{c}$, despite little
variations, for the on-site $s$-wave symmetry. Due to the
anisotropic SC gap for the mixed $s_{ex}+(d\pm id')$-wave paring
symmetry and the nodal SC gap for the mixed $(p\pm ip')+f$-wave
symmetry [see the dotted black lines in Fig.~\ref{fig3}(b)
and~\ref{fig3}(c)], the enhancement becomes more pronounced in a
$T^{\beta}$ power law below $T_{c}$ and $T^{\beta-1}$ at lower
temperature with $\beta=3$ for the mixed $s_{ex}+(d\pm id')$-wave
paring symmetry and $\beta=2$ for the the mixed $(p\pm ip')+f$-wave
symmetry, as denoted by the dotted lines in Figs. 4(e) and 4(f),
respectively.

\vspace*{.0cm}

In summary, we have provided a contrastive study on the kagome
superconductors at the van Hove filling with the incorporation of
the inequivalent TN bonds. Although the most favorable SC pairings
were derived from the electrons belonging to the same sublattice
sites, the consideration of the inequivalent TN bonds would result
in the SC pairings with multiple OAM components, and thus
contributed to the two-gap structures of the DOS. While the
spin-lattice relaxation exhibited distinct $T$ dependence in the SC
state for the three cases, a peak structure has been found for both
cases just below $T_{c}$. Unlike the coherent peak for the cases of
the on-site $s$- and the mixed $s_{ex}+(d\pm id')$-wave parings,
which was derived from both the enhancement of the SC DOS at the gap
edge and the non-zero SC coherent effect, the van Hove singularity
was crucial to the peak in the mixed $(p\pm ip')+f$-wave paring,
where only the enhancement of the SC DOS at the gap edge contributed
to the peak structure. In the vortex states, the cases for the
on-site $s$-wave and the mixed $s_{ex}+(d\pm id')$-wave parings
created discrete in-gap state peaks, which located on either side of
the zero energy. By contrast, the near-zero-energy and almost
dispersionless in-gap state peak occurred in the vortex core for the
case of the mixed $(p\pm ip')+f$-wave paring. Whereas the vortices
diminished the Hebel-Slichter (or Hebel-Slichter-like) peaks and
enhanced the $T$ dependence of $T^{-1}_{1}$ in the SC state for both
cases, the $T$ dependencies of $T^{-1}_{1}$ were also distinct with
respect to the different gap functions. While a more sophisticated
multi-orbital model is needed to make a direct comparison to the
experimental results, the SC pairing with multiple OAM components
originated from the inequivalent bonds on the kagome lattice and the
Hebel-Slichter (or Hebel-Slichter-like) peak of $T^{-1}_{1}$ below
$T_{c}$ due the vicinity to the van Hove filling were ecpected to
persist in a realistic multi-orbital description and should be
reflected in the experimental observations, provided that the system
situates close to the van Hove filling and the SC pairing is
nonlocal. The NMR experiments on different doping levels and on the
$T$ dependence of the $T^{-1}_{1}$ in the SC state both with and
without a perpendicular magnetic field were expected to testify the
theory.

\textit{note added.}---After completion of this study, we become
aware of recent interesting study on the vortex states in the kagome
superconductors by using the similar tight-binding
model~\cite{PDing1}. The SC vortex was simulated in
Ref.~\onlinecite{PDing1} by setting the spatial dependent pairing
amplitude $\Delta(\mathbf{r}_{i})=\Delta
\tanh(\frac{\mathbf{r}_{i}}{\xi})$, while the results in our study
were determined self-consistently.

\par The authors thank Professor Jian-Xin Li for fruitful discussions and valuable
suggestions. This work was supported by the National Natural Science
Foundation of China (Grant Nos. 11574069, 12074175) and the K. C.
Wong Magna Foundation in Ningbo University.

\end{document}